\documentclass[journal]{IEEEtran}
\usepackage{cite}
\usepackage{url}
\usepackage{hyperref}
\usepackage[utf8]{inputenc}
\usepackage{booktabs}
\usepackage{graphicx}
\usepackage{todonotes}

\begin{document}
\title{\texttt{regulAS}: A Bioinformatics Tool for the\\Integrative Analysis of Alternative Splicing\\Regulome using RNA-Seq data}
\author{\IEEEauthorblockN{Sofya Lipnitskaya\\}
\smallskip
\IEEEauthorblockA{Johannes Gutenberg University Mainz, Mainz, Germany\\}
\smallskip
\IEEEauthorblockA{solipnit@students.uni-mainz.de}
}
\date{July 16, 2023}

\maketitle
\smallskip

\begin{abstract}
The \texttt{regulAS} software package is a bioinformatics tool designed to support computational biology researchers in investigating regulatory mechanisms of splicing alterations through integrative analysis of large-scale RNA-Seq data from cancer and healthy human donors, characterized by TCGA and GTEx projects.
This technical report provides a comprehensive overview of \texttt{regulAS}, focusing on its core functionality, basic modules, experiment configuration, further extensibility and customisation.

The core functionality of \texttt{regulAS} enables the automation of computational experiments, efficient results storage and processing, and streamlined workflow management.
Integrated basic modules extend \texttt{regulAS} with features such as RNA-Seq data retrieval from the public multi-omics UCSC Xena data repository, predictive modeling and feature ranking capabilities using the \texttt{scikit-learn} package, and flexible reporting generation for analysing gene expression profiles and relevant modulations of alternative splicing aberrations across tissues and cancer types.
Experiment configuration is handled through \texttt{YAML} files with the \texttt{Hydra} and \texttt{\texttt{OmegaConf}} libraries, offering a user-friendly approach.
Additionally, \texttt{regulAS} allows for the development and integration of custom modules to handle specialized tasks.

In conclusion, \texttt{regulAS} provides an automated solution for alternative splicing and cancer biology studies, enhancing efficiency, reproducibility, and customization of experimental design, while the extensibility of the pipeline enables researchers to further tailor the software package to their specific needs.
Source code is available under the MIT license at \mbox{\url{https://github.com/slipnitskaya/regulAS}}.
\end{abstract}

\IEEEpeerreviewmaketitle

\section{Introduction}

Alternative splicing (AS) relates to a molecular mechanism that allows the generation of multiple mRNAs from a single gene to produce functionally distinct isoforms.
This process is largely regulated by RNA-binding proteins (RBPs) that control the recruitment of the splicing machinery defining which exons are included in the resulting transcripts.
Regulation of pre-mRNA splicing by RBPs is crucial for generating biological diversity in mammalian genomes, and this process is especially complicated in pathological conditions, such as cancer \cite{bonnal2020roles}.

The \texttt{regulAS} package allows easy and reliable exploration of the landscape of alternative splicing events and its candidate modulators across human tumor and healthy tissues through integrative transcriptomics analysis of large-scale RNA sequencing (RNA-Seq) datasets from diverse omics data sources, and by utilizing machine learning (ML) approach.

The purpose of this technical report is to provide a comprehensive overview and documentation of the \texttt{regulAS} software package for investigating alternative splicing regulome, utilizing external omics and associated phenotype datasets generated by The Cancer Genome Atlas (TCGA\footnote{\url{https://www.cancer.gov/tcga}}) and Genotype-Tissue Expression (GTEx\footnote{\url{https://www.gtexportal.org/home}}) projects.
Aimed at supporting researchers, \texttt{regulAS} offers a robust set of tools and functionalities to automate computational experiments, store, and process obtained results, and facilitate efficient workflows for alternative splicing research and cancer studies.

This report serves as a guide for both new and experienced users, offering detailed insights into the design principles, core features, and extensibility of \texttt{regulAS}.
By providing a high-level understanding of the software package, readers will gain the necessary knowledge to effectively utilize and harness the full potential of \texttt{regulAS} in their research endeavors.
The scope of this report encompasses a thorough exploration of the core functionality of \texttt{regulAS}, including its ability to automate computational experiments and handle result storage and processing.
Additionally, the report delves into the basic modules integrated into \texttt{regulAS}, such as data retrieval from external data sources, feature ranking capabilities, and the generation of tabular or visual summary output reports.

Furthermore, this report highlights the support for flexible experiment configuration offered by \texttt{regulAS} through the use of \texttt{YAML} files.
Lastly, the report addresses the extensibility aspect of \texttt{regulAS}, empowering end-users to expand its functionality beyond the provided modules.
By enabling the development of custom modules in Python, \texttt{regulAS} offers researchers the freedom to integrate their specialized algorithms, models, or data processing techniques seamlessly.

\begin{figure}
  \includegraphics[width=\linewidth]{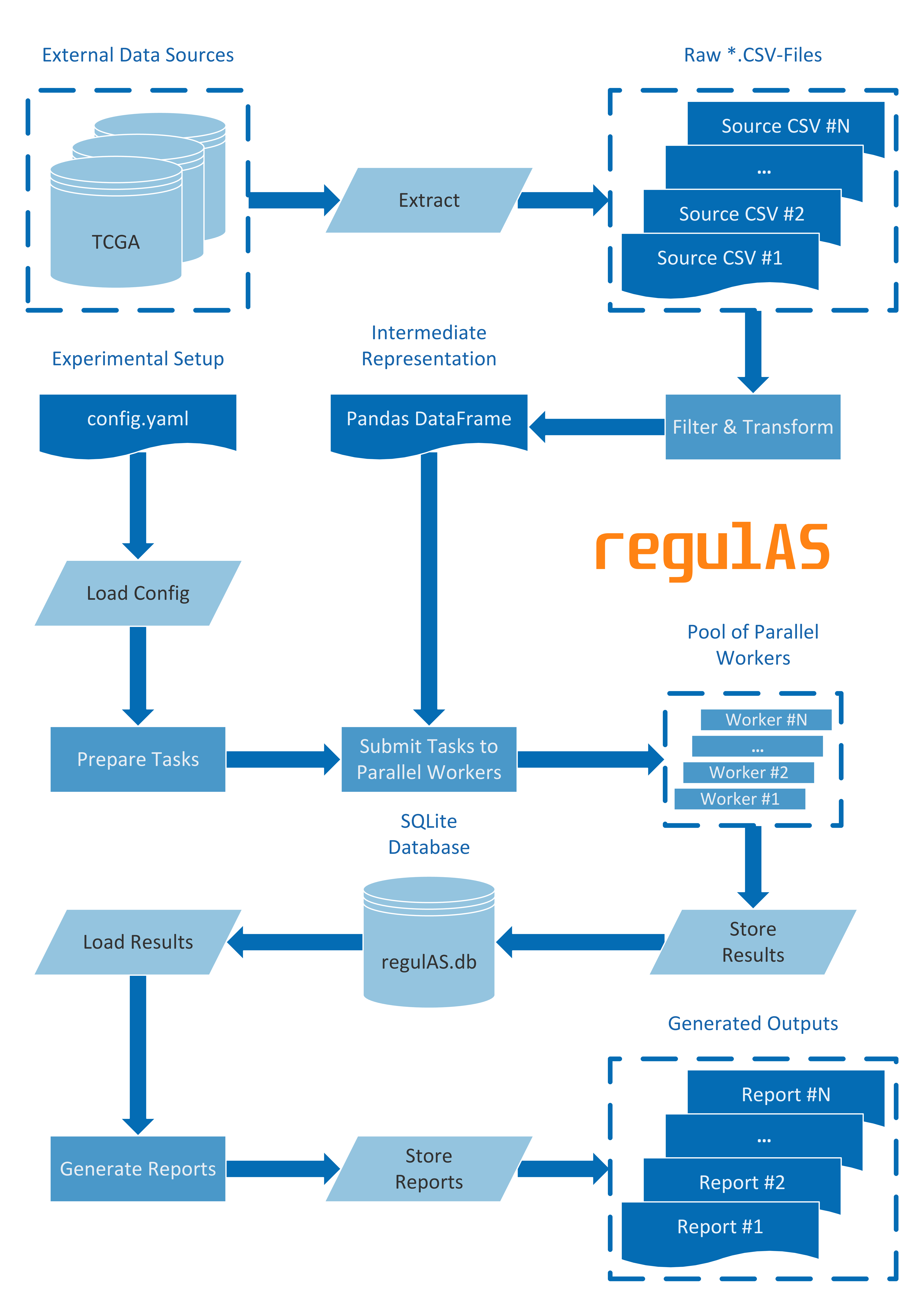}
  \caption{
  Workflow of the \texttt{regulAS} bioinformatics package for integrative transcriptomic analysis of alternative splicing events based on machine learning approach and RNA-Seq data from TCGA and GTEx data sources.
  }
  \label{fig:workflow}
\end{figure}

\subsection{Problem Definition}

In the field of computational biology and bioinformatics, researchers often face numerous challenges when integrating and analysing large-scale multi-omics datasets based on various public data sources \cite{picard2021integration}.
These challenges include the need for efficient experimental design and automation, effective management and processing of results, seamless integration with external databases, and the ability to generate informative reports for downstream analyses and interpretation purposes to support the findings and experimental results.

Traditionally, researchers have had to rely on manual execution of experiments, leading to time-consuming and error-prone processes.
Additionally, the management and processing of the vast amount of data generated from these experiments pose significant difficulties, often requiring extensive manual effort and specialized software tools.
Furthermore, integrating with external multi-omics data sources and retrieving relevant datasets for analysing gene expression profiles and alternative splicing patterns across different conditions (e.g., tumor and/or healthy), cancer types (e.g., primary tumor and/or tumor-adjacent tissues) appear to be challenging and time-consuming, also commonly necessitating specialized knowledge and technical skills.

Other challenges are associated with the raw data processing and sample aggregation, both of which are essential for effective subsequent analysis---such as feature importance assessment or predictive survival analysis---and interpretation of the results (e.g., for studying gene expression abnormalities across tissues and physiological conditions).

Additionally, comprehensive summary reports and visualizations that present research findings in a clear and concise manner are necessary for effective communication and collaboration.
Thus, researchers need standardized computational solutions for integrative omics analysis to streamline data exploration and computational workflows towards data-informed decision-making about downstream experiments,  hence contributing to a better understanding of biological systems.

\subsection{Design Principles}

The design of \texttt{regulAS} follows a few key concepts.
First, a low barrier to entry was prioritized, which implied following the ``low-code'' paradigm.
Specifically, the base modules allow performing computational experiments by describing the desired workflow in a form of \texttt{YAML} configuration files, without writing code in Python.

Second, the consistency among external and internal interfaces, and intermediate representations was addressed, which required considerable efforts in designing the architecture, however, paid off in an easy extensibility of \texttt{regulAS} to user-defined data loaders, ML-models, metric evaluators, and report generators.

Third, as the use of the \texttt{scikit-learn} \cite{sklearn2011pedregosa} package has become a \textit{de facto} standard for performing computational experiments in a Python-based environment, \texttt{regulAS} focuses on compatibility with the \texttt{scikit-learn} model API.
As a part of the experimental workflow, \texttt{regulAS} utilizes a supervised ML-based approach for predictive modelling and feature ranking tasks to identify relevant candidate RBPs of AS changes based on RNA-Seq data of gene expression profiles of prospective RNA splicing modulators and matching junction reads data to reflect exon-skipping events \cite{schafer2015alternative} for genes of interest.

Fourth, the base workflow steps such as data acquisition and preparation, model training and evaluation, and generation of reports were isolated---as depicted in \autoref{fig:workflow}---to preserve modularity (and, therefore, extensibility) and support reuse of modules in the experiments.
The isolation of base steps is accomplished by keeping the intermediate results in a persistent storage, namely a relational database built upon the SQLite engine \cite{sqlite2023hipp}.

Finally, because bioinformatics and computational biology are rapidly evolving fields, the specific methods and approaches may become outdated quite fast.
To address this, the architecture of \texttt{regulAS} was designed to be \textit{modular} and \textit{extensible}, to allow practitioners to leverage existing libraries and experimenting with their own components.
For that, the implementation of \texttt{regulAS} strived for a clean, readable, and consistent code.

\section{Package Organization}

As the main priority in the design was ease of use, \texttt{regulAS} follows a flat package structure and relies on the NumPy \cite{harris2020array} and Pandas \cite{mckinney2010pandas} types and data structures.

The extensibility of \texttt{regulAS} through custom modules provides researchers with unparalleled flexibility in adapting the software package to their unique research needs.
By leveraging the Python programming language, researchers can harness the rich ecosystem of libraries and tools to develop custom modules that seamlessly integrate with \texttt{regulAS}'s core functionality.

Custom modules can be easily integrated into \texttt{regulAS}, utilizing its modular architecture and well-defined extension points.
This allows researchers to leverage the existing functionalities of \texttt{regulAS} while incorporating their own custom logic and algorithms.
The integration process ensures a smooth and cohesive workflow, enabling researchers to exploit the full potential of \texttt{regulAS} while building upon its foundations.

\begin{figure*}
  \includegraphics[width=\linewidth]{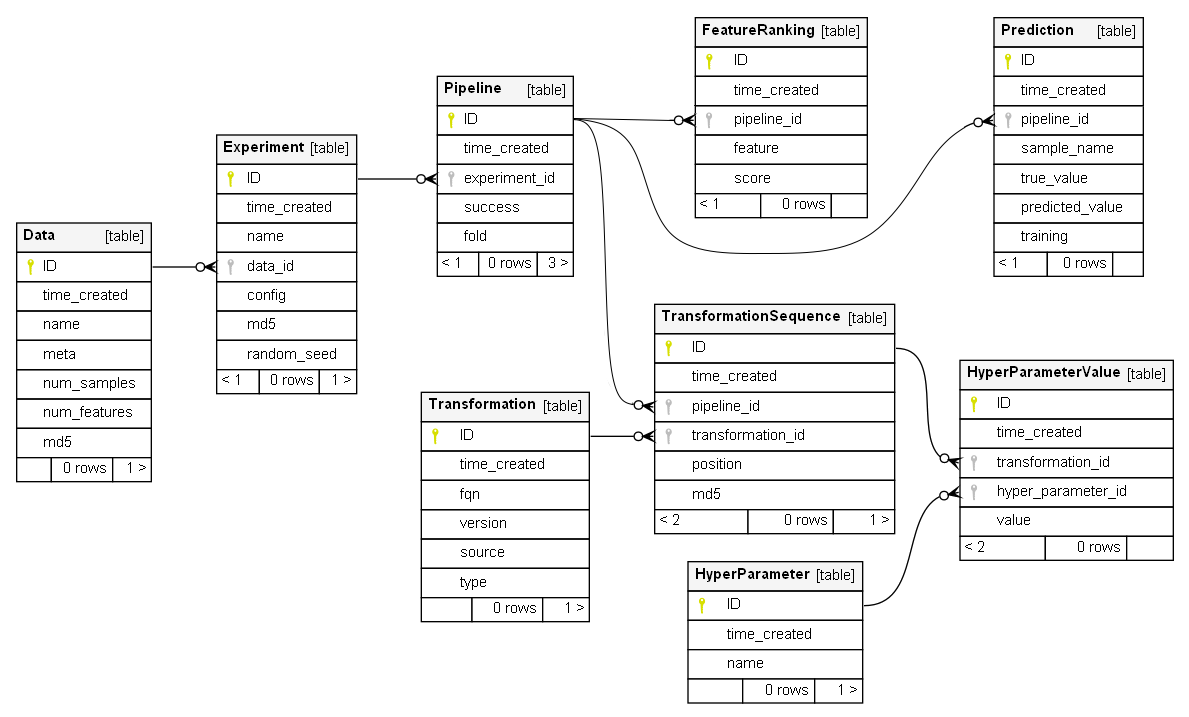}
  \caption{Schema of the \texttt{regulAS} database for storing experimental setups and results.}
  \label{fig:database}
\end{figure*}

\subsection{Core Functionality}

\texttt{regulAS} provides researchers with the ability to automate computational experiments, reducing the manual effort and time required for experiment execution. By defining experiment configurations and parameters, researchers can easily set up and execute complex computational experiments with ease. This automation feature allows for increased productivity, reproducibility, and scalability in research workflows.

\texttt{regulAS} offers a flexible and user-friendly approach to experiment configuration through the use of \texttt{YAML} (YAML Ain't Markup Language) files. Experiment configuration is facilitated by the \texttt{Hydra} library \cite{yadan2019hydra}, developed by Facebook Research, and the underlying \texttt{OmegaConf} library, which provides a powerful configuration system.

\texttt{Hydra}, a popular open-source Python library released by Facebook Research, is utilized in \texttt{regulAS} for managing experiment configuration. \texttt{Hydra} simplifies the process of working with \texttt{YAML} files by providing a hierarchical and modular configuration system. It allows researchers to easily define and organize their experiment parameters, enabling the configuration of various components and options.

Underlying \texttt{Hydra}, \texttt{OmegaConf} provides the core configuration capabilities. It enables researchers to read and merge \texttt{YAML} files, override default configurations, and access experiment parameters programmatically. \texttt{OmegaConf} also supports interpolation, which allows referencing and reusing values across the configuration, enhancing flexibility and reusability.

As shown in \autoref{fig:workflow}, after successfully loading the experiment configuration file, \texttt{regulAS} loads the data, performs the training/testing split, and stores the experimental setup in its database.
Next, based on the configuration file and data splits, \texttt{regulAS} extracts individual tasks and submits them asynchronously to a pool of pre-allocated Python \texttt{multiprocessing} workers.

Then, each worker returns a task identifier, a success flag and the corresponding output or \texttt{None} if the run failed.
For a successful run, \texttt{regulAS} stores the obtained predictions and feature importance scores (if available) into the database.

\subsection{Data Loading}

\texttt{regulAS} is bundled with base data loading modules that encompass retrieval from both local (Python \texttt{pickle}) and remote (UCSC Xena\footnote{\url{https://xena.ucsc.edu/}} data hubs) data sources \cite{goldman2020visualizing}.
The latter data loader acquires the required information from the remote databases, performs filtering and transformation if needed, and stores the ready-to-process \texttt{pickle}-serialized data locally.
Respectively, the local data loader allows deserializing the downloaded data and feeding them into the downstream processing pipeline.

Researchers can also develop custom modules to handle specialized data loading tasks within \texttt{regulAS}.
These modules can be designed to integrate with specific data sources, formats, or APIs, allowing researchers to seamlessly import and preprocess their unique datasets.
By extending \texttt{regulAS} with custom data loading modules, researchers gain the flexibility to incorporate their proprietary or specialized data sources into their computational experiments, ensuring compatibility and efficiency.

\subsection{Persistence}

\texttt{regulAS} offers efficient storage and processing of obtained results, ensuring that researchers can effectively manage and analyze their data.
One of the key components of the result storage in \texttt{regulAS} is the utilization of the SQLite database engine.
This lightweight and fast database engine enables reliable storage and retrieval of experiment results, providing researchers with a robust and scalable solution for data management.
With the built-in capabilities for querying and accessing stored results, researchers can efficiently explore and extract valuable insights from their data.
The visual representation of the database schema is depicted in \autoref{fig:database}.

Table \texttt{Data} represents the root entity of the database.
An \texttt{Experiment} depends on \texttt{Data} and, in turn, encapsulates several machine learning \texttt{Pipeline}s.

Each \texttt{Pipeline} is associated with a number of data transformation steps (\texttt{TransformationSequence}), and every step is defined by the \texttt{Transformation} itself, \texttt{HyperParameter}s used, and their corresponding values (\texttt{HyperParameterValue}).

Finally, a successfully finished \texttt{Pipeline} yields a list of model \texttt{Prediction}s equipped with the corresponding \textit{true} values, and---in case the underlying ML-model (\texttt{Transformation}) supports it---numerical scores for \texttt{FeatureRanking}.

\subsection{Report Generation}

The reporting module encompasses both intermediate and final report generation.
The former type of report focuses on the assessment of performance of an ML-model and scoring of the feature importance, while the latter is responsible for producing documents for end-users and includes tabular (CSV) and visual (bar graph and scatter plot) reports.

By design, the report configuration organizes the report steps in a directed acyclic graph, those leaf nodes are expected to produce (although not required) tabular or visual documents.
In turn, each leaf node depends on some preceding nodes, which are to produce intermediate reports and not required to store any documents.

When loading the report configuration file, \texttt{regulAS} checks for missing and circular dependencies.
In case of any, the report generation interrupts, and a corresponding error appears in the standard output.
If there is no cycle in dependencies, and all of them are satisfied, \texttt{regulAS} organizes the reporting steps into a linear structure that is processed sequentially, starting from the independent reports.

The built-in report generation module enables \texttt{regulAS} to produce model evaluation and feature ranking in a tabular form (implemented on top of \texttt{pandas.DataFrame}).
For the visual reports, bar graphs are available for the performance evaluation and feature ranking purposes, while scatter plots allow visualizing correlations.

Additionally, custom modules can be developed to expand \texttt{regulAS}'s report generation capabilities.
Researchers can design and implement modules that generate reports in specific formats, layouts, or styles, aligning with their specific reporting needs.
These custom report generation modules can produce reports that include additional visualizations, statistical summaries, or customized annotations, providing researchers with comprehensive and tailored reports for their analyses and presentations.

\section{Conclusion}

This document provides a brief yet comprehensive summary of \texttt{regulAS}, a software package that addresses the needs of researchers in computational biology and bioinformatics.
By automating computational experiments, managing results, and providing flexible configuration options, \texttt{regulAS} simplifies research workflows and enhances productivity.
The integration of basic modules and the ability to extend \texttt{regulAS} with custom modules further expand its capabilities, ensuring that researchers can leverage the software package to its full potential for their unique research objectives.
The project is under active development, and the future work includes performance improvements and extended functionality of the experiment management.

\subsection{Citing \texttt{regulAS}}

When using \texttt{regulAS} in academic work, authors are encouraged to reference this work via Zenodo\footnote{\url{https://doi.org/10.5281/zenodo.8152782}} or by following the instructions on the GitHub repository\footnote{\url{https://github.com/slipnitskaya/regulAS}}.

\bibliographystyle{ieeetr}
\bibliography{report}

\end{document}